\documentclass[fleqn]{llncs}

\usepackage{times}  % any effect?
\usepackage[all]{xy}
\usepackage{enumitem}
\renewcommand{\tt}{\rm \ttfamily \bfseries }

\usepackage{amssymb}
\usepackage{amsmath}
\usepackage{mathtools}
\usepackage{mdframed}
\usepackage{url}

\usepackage{xspace}
\usepackage[T1]{fontenc}
\usepackage[mathlines]{lineno}

\usepackage{graphics}
\usepackage{xcolor}

\newcommand{\COMMENT}[1]{}

\newcommand{\w}{\noindent{\rm \phantom{XX}}} % tab

\newcommand{\Root}{\ensuremath{\mathcal{R}}}
\newcommand{\nodes}{\ensuremath{\mathcal{N}}}

\newcommand{\xlabel}{\ensuremath{\mathcal{L}}}
\newcommand{\xsucc}{\ensuremath{\mathcal{S}}}
\newcommand{\dom}{\ensuremath{\mathcal{D}}}
\newcommand{\Path}{\ensuremath{\mathcal{P}}}
\newcommand{\ren}{\Theta}

\newcommand{\vars}{\ensuremath{\mathcal{X}}}

\renewcommand{\tt}{\rm \ttfamily \bfseries}
\newcommand{\codefont}{\small\tt}
\newcommand{\code}[1]{\mbox{\codefont{#1}}}
\newcommand{\ccode}[1]{``\code{#1}''}
\newcommand{\us}{\raise-.8ex\hbox{-}}
\newcommand{\power}{\char94} % caret
\newcommand{\xtilde}{\!\raise-.75ex\hbox{\char`\~}} % low tilde

\newcommand{\equprogram}[1]{%
\def\separator{1.03333ex}%
\frenchspacing%
\refstepcounter{equation}%
\par\vspace\separator\hspace{.5em}%
\scalebox{0.95}{$\vcenter{\codefont\noindent{#1}}$}%
\raisebox{0.3ex}{\kern-.5em\llap{\rm (\theequation)}}
\par\vspace\separator\noindent\kern-.0em%
}

\begin{document}
% remove for proceedings
% \pagestyle{plain}
\sloppy
% \linenumbers

\title{Making Bubbling Practical}

\author{
Sergio Antoy\thanks{Supported in part by NSF grant 1317249.}
\kern1em
Steven Libby
}
\institute{
Computer Science Dept., Portland State University, Oregon, U.S.A.\\
\email{\{antoy,slibby\}@cs.pdx.edu}\\[1ex]
}

\maketitle

\begin{abstract}
Bubbling is a run-time graph transformation studied for the execution of non-deterministic steps in functional logic computations. This transformation has been proven correct, but as currently formulated it requires information about the entire context of a step, even when the step affects only a handful of nodes. Therefore, despite some advantages, it does not appear to be competitive with approaches that require only localized information, such as backtracking and pull-tabbing. We propose a novel algorithm that executes bubbling steps accessing only local information. To this aim, we define graphs that have an additional attribute, a dominator of each node, and we maintain this attribute when a rewrite and/or bubbling step is executed. When a bubbling step is executed, the dominator is available at no cost, and only local information is accessed. Our work makes bubbling practical, and theoretically competitive, for implementing non-determinism in functional logic computations.
\end{abstract}


%%%%%%%%%%%%%%%%%%%%%%%%%%%%%%%%%%%%%%%%%%%%%%%%%%%%%%%%%%%%%%%%%%%%%%%%%%
\section{Introduction}
\label{Introduction}

Consider, in Curry's syntax  \cite{Hanus16Curry},
a function for computing the body mass index
\cite{wiki:bmi} of an individual:
\equprogram{%
  bmi x = weight x / (height x) \power{} 2
}
In a functional logic language,
for example, to find someone with a bmi greater than 25,
we evaluate (solve) the following disequation:
\equprogram{%
  \label{bmi-equation}%
  bmi $t$ > 25
}
where $t$ is a non-deterministic choice among the members
of some set of interest.
For example, for the sake of completeness and simplicity let us assume
that the set of interest consists of \emph{Alice} and
the parents of \emph{Bob}, i.e., $t =\,\,$\emph{Alice \code{?} parent Bob}.
Weights, heights and parents of individuals are defined by suitable functions.

The expressions manipulated by a program are abstracted by graphs.
Fig.~\ref{bmi-graph} shows one of these graphs, a state of
the computation of disequation (\ref{bmi-equation}).
\begin{figure}[hbt]
  \begin{mdframed}
    \centerline{
      \xymatrix@C=-2pt@R=15pt@C=5pt{
        & & & \code{>} \ar@{-}[dll] \ar@{-}[drr] \\
        & \code{/} \ar@{-}[dl] \ar@{-}[dr] & & & & \code{25} \\
        \code{weight} \ar@{-}@/_9pt/[ddr]
               & & \code{\power} \ar@{-}[dl] \ar@{-}[dr] \\
        & \code{height} \ar@{-}[d] & & \code{2} \\
        & \code{?} \ar@{-}[dl] \ar@{-}[dr] \\
        \code{Alice} & & \code{parent} \ar@{-}[d] \\
        & & \code{Bob}
      }
    }
    \caption{
      \label{bmi-graph}
      Graphical representation of a state of the computation of
      disequation (\ref{bmi-equation}) for
      $t=\emph{Alice}\, \code{?}\, \emph{parent~Bob}$.
    }
  \end{mdframed}  
\end{figure}
We draw the attention on some elements of this expression/graph,
hereafter referred to as $e$, that are relevant to our discussion.
The subexpression of $e$ rooted by ``\code{?}'' is a
non-deterministic choice, or more simply a \emph{choice}.
Its values are the values of either of its alternatives.
The rest of $e$ is the \emph{context} of the \emph{choice}.
The node of $e$ labeled by ``\code{/}'' is a \emph{dominator}
of the \emph{choice}, i.e., any path from the root of $e$
to the \emph{choice} goes through the dominator.

The focus of this paper is how to compute the values of
an expression of this kind.
To obtain all the values of the expression,
all the alternatives of the \emph{choice} must be considered.
Each alternative is evaluated within the context
of the \emph{choice}.
Since the computation for a selected alternative ``consumes'' the context,
we need a fresh/new context for each alternative.

Historically, three main techniques have been proposed
for this problem.  Backtracking \cite{knuth} first selects
an alternative of the  \emph{choice} and computes the entire
expression with this alternative, then it rollbacks
the computation to recreate the context before selecting the other alternative.
By contrast,
\emph{pull-tabbing} \cite{Antoy11ICLP,Brassel2011PHD,BrasselHuchAPLAS07}
and \emph{bubbling} \cite{AntoyBrownChiang06Termgraph,AntoyBrownChiang06RTA}
``clone'', though in different ways, the context
for each alternative, so that each alternative
can be computed in its own context.
Our work focuses on bubbling.
Loosely speaking, a bubbling step ``swaps'' a \emph{choice} with
some dominator of the choice and in the process clones
the portion of the graph between dominator and \emph{choice}
to provide a context to each alternative of the \emph{choice}.
A formal definition and a detailed example will be provided later.

Bubbling, as currently formulated, has some advantage over the other techniques,
but also has a fatal flaw: the
necessity of traversing the entire expression to find
the immediate dominator of the \emph{choice}
anytime a bubbling step is executed.
In this paper, we describe a novel technique that stores some
additional information in a graph, and maintains it
during a computation, so that when a bubbling step
is executed, a dominator, which may not be immediate
but is still viable, is readily available
without traversing the graph.

This paper is organized as follows.
Section \ref{Curry:sec} is a mini-introduction to Functional Logic
Programming and Curry.
Section \ref{Background} provides background information about
the kind of programs our contribution applies to,
the definition of graphs which slightly extends
the standard one with additional information in the nodes,
the kind of steps allowed in a computation,
and the definition of bubbling which also extends the standard one
with additional information.
Section \ref{Algorithms} presents two original algorithms,
one to execute a bubbling step using only local information,
the other to execute a rewrite step that preserves
information used by the previous algorithm.
Section \ref{Correctness} states the correctness of the
original algorithms, but without formal proofs.
Only an informal argument is provided that hopefully
helps understanding the inner working of the algorithms.
Section \ref{Related Work} summarizes work in this area
and Section \ref{Conclusion} offers our conclusion.

\section{Functional Logic Programming and Curry}
\label{Curry:sec}

We briefly recall elements of functional logic languages
and Curry that may help understanding our contribution.
More details can be found in surveys \cite{AntoyHanus10CACM,Hanus13}
and the language report \cite{Hanus16Curry}.

Curry is a declarative multi-paradigm language
combining functional and logic programming
with a syntax close to Haskell's \cite{PeytonJones03}.
A Curry program declares constructor symbols via data declarations
and operation (or function) symbols via defining rules.
The following example shows  these elements:
\equprogram{%
  data Bintree = Leaf | Branch Int Bintree Bintree \\
  isin \us{} Leaft = False \\
  isin n (Branch i left right) \\
  \w\w = n == i || isin n left || isin n right 
}
\emph{Leaf} and \emph{Branch} are the constructors of a type
binary tree of integers and \emph{isin} is an operation telling whether
an integer is in a tree.  The operators \ccode{==} and \ccode{||},
which stand for Boolean equality and conjunction,
are defined in a standard library.
A function may be applied to a logic (or free) variable
which is instantiated non-deterministically by narrowing,
if needed.
For example, the evaluation of:
\equprogram{%
  isin 5 x where x free
}
binds $x$ to \emph{Leaf} and \emph{Branch u v w} non-deterministically,
where $u$, $v$ and $w$ are free variables as well.

Non-determinism originates from free variables, as shown above,
and overlapping rules, the epitome of which
are those of the \emph{choice} operation:
\equprogram{%
\label{choice-rules}%
x ? \us{}  =  x \\
\us{} ? y  =  y
}
The textual order of rules is irrelevant.
Each rules of the \emph{choice} is equally applicable.
Thus, the expression $(0\, \code{?}\, 1)$ evaluates to $0$ and $1$
with the value non-deterministically chosen.

Curry has a variety of other syntactic and semantic features 
usually found in a general-purpose programming language.
These features are not needed to understand our contribution
and we only mention a few ones for the sake of completeness:
the compile-time definition of infix binary operators
with user-defined precedence and associativity;
higher-order functions,
but without higher-order narrowing;
declarative input/output in the monadic style;
set functions for encapsulated non-determinism;
and functional patterns.
Curry programs can be partitioned into modules for programming
in the large and they can invoke external functions, i.e.,
functions that are not coded in Curry, for interacting with
the operating system.

The mainstream compiler/interpreter of 
Curry, {\sc Pakcs} \cite{Hanus17PAKCS}, provides
an extensive set of libraries,  including primitives for
building graphical user interfaces, interactive web pages,
distributed applications and accessing database engines.

\section{Background}
\label{Background}

The theory of \emph{graph rewriting}
\cite{EchahedJanodet97IMAG,Plump99Handbook} is heavy.
It would be impossible to review here
even the simplest and most fundamental concepts.
Therefore, we only recall those elements of the theory,
in particular their notations,
that are necessary to present our contribution.

The core representation of a program
in current Curry compilers
\cite{AntoyJost16LOPSTR,BrasselHanusPeemoellerReck11,Hanus17PAKCS}
is a \emph{Limited Overlapping Inductively Sequential},
abbreviated \emph{LOIS}, graph rewriting system.
In \emph{LOIS} systems, the rules
are left-linear and constructor-based.
The left-hand sides of the rules are organized
in a hierarchical structure called a 
\emph{definitional tree} \cite{Antoy92ALP}.
In \emph{LOIS} systems, the only operation
with overlapping rules is the \emph{choice}
defined in (\ref{choice-rules}).

We ignore \emph{free} (also called \emph{logic})
variables in the rest of our discussion
since they can be replaced by \emph{generators} \cite{AntoyHanus06ICLP}.
Thus, narrowing in a \emph{LOIS} system with free variables is equivalent
to rewriting in a similar \emph{LOIS} system without free variables.

Non-determinism in functional logic programming
comes from both free variables and overlapping rules.
Since free variables are banned and \emph{LOIS} systems
allow a single overlapping rule,
the \emph{choice} operation is the only source
of non-determinism in our programs.
Bubbling handles the evaluation of expression
with occurrences of the \emph{choice} operation,
hence all the non-determinism of our programs.

\subsection{Graphs}

A graph is the formalization of the intuitive and often informal
notion of expression.
Our definition of a graph is a minor extension of~\cite{EchahedJanodet97IMAG}
from which we entirely adopt notations and terminology.
We add a new attribute,
$\dom$, intended to map every non-root node, $n$, of a graph
to some proper dominator of $n$.
We will say that the $\dom$ attribute of a graph $g$ is \emph{correct}
iff for every node $n$ of $g$, $\dom_g(n)$ is indeed a dominator of $n$.
We recall that given a graph $g$ and two nodes $d$ and $n$,
$d$ \emph{dominates} $n$ iff
every path from the root of $g$ to $n$ contains $d$.
It follows that every node trivially dominates itself.
Throughout this paper, when we say that $d$ dominates $n$,
or is a dominator of $n$,
we mean \emph{properly}, i.e., $d \ne n$, unless
when explicitly stated otherwise.
A consequence is that a dominated node is never the root of a graph.
A node may have many dominators, in particular, the root of
a graph $g$ dominates every other node of $g$.
Typically, $\dom_g(n)$ will be the immediate (closest) dominator of $n$,
but we do not enforce this condition because it is convenient to
relax it in some situations.
We will see later that a bubbling step clones the nodes
in a path from a dominator node $d$
to a dominated node $n$.
Therefore, the closer $d$ is to $n$,
the fewer nodes are cloned in a bubbling step at $n$.
The bubbling algorithm also uses
the \emph{predecessor} relation, which
is the inverse of the \emph{successor} relation,
hence we do not explicitly define a predecessor attribute.
Occasionally we will equate a node $n$ of a graph $g$
with the subgraph of $g$ rooted by $n$ since they are in a bijection.

% \todo{Are expressions admissible graphs, dags only?}

\begin{definition}[Expression]
\label{def:expression}
Let $\Sigma$ be a \emph{signature},
% partitioned into \emph{constructor} and \emph{operation} symbols,
$\vars$ a countable set of \emph{variables},
$\nodes$ a countable set of \emph{nodes}.
A \emph{graph} or \emph{expression} over
$\langle\Sigma,\nodes,\vars\rangle$
is a 5-tuple 
$g=\langle \nodes_g,\xlabel_g,\xsucc_g,\Root_g,\dom_g\rangle$
such that: 
\begin{enumerate}
\renewcommand{\labelenumi}{\arabic{enumi}.}
\item {}
$\nodes_g \subset \nodes$ is the set of nodes of $g$;
\item {}
$\xlabel_g : \nodes_g \to \Sigma \cup \vars$ is the \emph{labeling}
function mapping each node of $g$ to a signature symbol or a variable;
\item {}
$\xsucc_g : \nodes_g \to \nodes_g^*$
is the \emph{successor}
function mapping each node of $g$ to a possibly empty string of
nodes of $g$ such that
if $\xlabel_g(n)=s$, where $s \in \Sigma \cup \vars$,
and (for the following condition, we assume that a variable has arity zero)
$\mathit{arity}(s)=k$,
then there exist $n_1,\ldots,n_k$ in $\nodes_g$ such that
$\xsucc_g(n)=n_1 \ldots n_k$;
\item{}
$\Root_g \in \nodes_g$ is a distinguished node of $g$
called the \emph{root} of $g$;
\item {}
$\dom_g : \nodes_g - \{ \Root_g \} \to \nodes_g$
  is the \emph{dominator} function
  mapping every non-root node $n$ of $g$ to some node $g$ such that
  $\dom_g(n)$ is a dominator of $n$;
\item{}
if $\xlabel_g(n_1) \in \vars$ and $\xlabel_g(n_2) \in \vars$
and $\xlabel_g(n_1) = \xlabel_g(n_2)$, then $n_1 = n_2$,
i.e., every variable of $g$ labels one and only one node of $g$; and
\item {}
for each $n \in \nodes_g$, either $n=\Root_g$
or there is a path from $\Root_g$ to $n$,
i.e., every node of $g$ is reachable
from the root of $g$.
\end{enumerate}
\end{definition}
Typically we say ``expression'' when talking about programs
and ``graph'' when making formal claims.
All our graphs are ``term graphs'' in the terminology of
\cite{EchahedJanodet97IMAG}, i.e., they have exactly one root.
Expressions have both a textual and a graphical notation.
The textual notation is in Curry's syntax \cite{Hanus16Curry}.
The notation presents node labels rather than the nodes themselves.
Function application defines the \emph{successor} relation,
e.g., in $f\,t_1 \ldots t_n$, the root of $t_i$ is a
successor of the root of the entire expression which is labeled by $f$.
A \emph{where} clause introduces local declarations
including variables.  ``Variable'' is an overloaded concept in Curry.
There are \emph{bound} variables, introduced in the left-hand side of a rule,
\emph{logic} variables, introduced by a \emph{free} declaration,
and \emph{local} variables, introduced by \emph{let}
or \emph{where} constructs.
A principal use of these constructs is for sharing nodes.
For example, in the graph of Fig.~\ref{bmi-graph}
the node labeled by the \emph{choice} is the successor of
two distinct nodes that ``share'' it.
The textual representation of the graph is:
\equprogram{%
  weight x / (height x) \char94{} 2 > 25 \\
  \w\w where x = Alice ? parent Bob
}
The ``where'' clause introduces the local variable $x$.
The variable identifies a node.
The binding of the variable defines the subexpression rooted by this node.
Node identifiers are arbitrary and irrelevant to most purposes.
In fact, graphs that differ only
for a renaming of nodes \cite[Def.~15]{EchahedJanodet97IMAG}
are considered equal.

\subsection{Computations}
\label{Computations}

Sections 2 and 3 of \cite{EchahedJanodet97IMAG} 
formalize key concepts of graph rewriting such as
\emph{replacement}, \emph{matching}, \emph{homomorphism},
\emph{rewrite rule}, \emph{redex}, and \emph{step}
in a form ideal for our discussion.
Therefore, we adopt these definitions in their entirety,
including their notations.
Similar treatments are discussed
in \cite{BaaderNipkow98,Terese03,Plump99Handbook}.
All these treatments do not include
the \emph{dominator} attribute, which is specific to our work.
Hence, after some preliminaries,
we will formally describe how the \emph{dominator} attribute
is maintained throughout a computation.

A \emph{computation} of an expression $e$ in a program $P$ is a sequence 
$e=e_0 \to e_1 \to \ldots$ such that $e_i \to e_{i+1}$ is a
rewrite \emph{step} according to a rule of $P$.
Later in this paper we will also
allow a second kind of step called bubbling \cite{AntoyBrownChiang06Termgraph,AntoyBrownChiang06RTA}
which is independent of the rules of a program.
``\emph{Evaluation}'' is a synonym of a computation in which the intent
is to find a \emph{value} (constructor normal form) of an expression.
A \emph{rewrite step} is the replacement in a graph of an instance
of a rewrite rule's left-hand side (the \emph{redex}) with the corresponding
instance of the rule's right-hand side (the \emph{contractum} or replacement).
If $s$ is a subexpression of an expression $e$,
the \emph{context} of $s$ in $e$ is the portion of $e$ disjoint from $s$.
\emph{Choice} reductions, which are non-deterministic steps, are limited
to the root of an expression where a \emph{choice} has an empty context.
This makes it easy to evaluate both alternatives concurrently and independently.
By contrast, reducing a \emph{choice} with a non-empty context
entails an irrevocable commitment, since there is a single context
for two alternatives
Bubbling steps
are equivalent to non-deterministic steps in the sense
that they have the potential to produce exactly the intended results
of an expression \cite{AntoyBrownChiang06RTA},
but without any irrevocable commitment to either alternative.

\subsection{Bubbling}
\label{Bubbling}

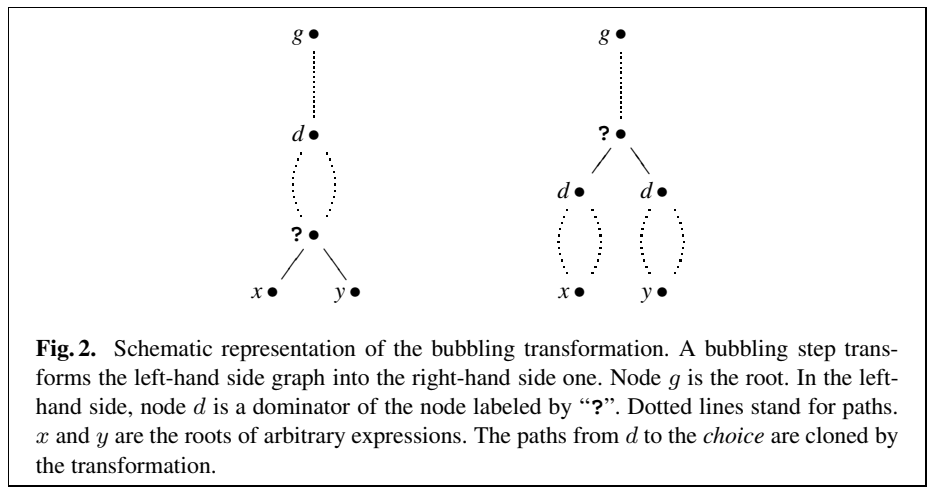
\begin{figure}[hbt]
  \begin{mdframed}
    \centerline{\it
      \xymatrix@C=-2pt@R=10pt@C=5pt{
        & \llap{g\,}\bullet \ar@{.}[dd] \\
        \\
        & \llap{d\,}\bullet \ar@{.}@/_8pt/[dd] \ar@{.}@/^8pt/[dd] \\
        \\
        & \llap{\code{?}\,}\bullet \ar@{-}[ld] \ar@{-}[rd]  \\
        \llap{x\,}\bullet & & \llap{y\,}\bullet
      }
      \hspace*{8em}
      \xymatrix@C=-2pt@R=10pt@C=5pt{
        & \llap{g\,}\bullet \ar@{.}[dd] \\
        \\
        & \llap{\code{?}\,}\bullet \ar@{-}[ld] \ar@{-}[rd]  \\
        \llap{d\,}\bullet \ar@{.}@/_8pt/[dd] \ar@{.}@/^8pt/[dd] & &
        \llap{d\,}\bullet \ar@{.}@/_8pt/[dd] \ar@{.}@/^8pt/[dd] \\
        \\
        \llap{x\,}\bullet & & \llap{y\,}\bullet
      }
    }
    \caption{
      \label{bubbling-pic}
      Schematic representation of the bubbling transformation.
      A bubbling step transforms the left-hand side graph
      into the right-hand side one.
      Node $g$ is the root. 
      In the left-hand side, 
      node $d$ is a dominator of the node labeled by \ccode{?}.
      Dotted lines stand for paths. $x$ and $y$ are the roots of
      arbitrary expressions.
      The paths from $d$ to the \emph{choice} are cloned by the transformation.
    }
  \end{mdframed}
\end{figure}

Fig.~\ref{bubbling-pic} informally shows a bubbling step.
The step ``moves'' a node labeled by the \emph{choice} up to a dominator
and clones the paths from the dominator to the choice 
for each alternative of the choice.
We begin with an auxiliary concept.
\begin{definition}[Partial renaming]
\label{def:renaming}
Let $g=\langle \nodes_g,\xlabel_g,\xsucc_g,\Root_g,\dom_g\rangle$
be a term graph over $\langle\Sigma,\nodes,\vars\rangle$,
$\nodes_p$ a subset of $\nodes_g$ and
$\nodes_q$ a set of nodes disjoint from $\nodes_g$.
A \emph{partial renaming} of $g$ with respect to $\nodes_p$ and $\nodes_q$
is a bijection $\ren:\nodes \to \nodes$
such that:
\begin{equation}
\label{eq:renaming}
\ren(n)=\left\{
\renewcommand{\arraystretch}{1.1}
  \begin{array}{@{} l l @{}}
    n' \hspace*{1em} & \mbox{where $n' \in \nodes_q$, 
                             if $n \in \nodes_p$;} \\
    n & \mbox{otherwise.}
  \end{array}
\right.
\end{equation}
Similar to substitutions, we call $\nodes_p$ and $\nodes_q$,
the \emph{domain} and \emph{image} of $\ren$, respectively.
We overload $\ren$ to graphs as follows:
$\ren(g)=g'$ is a graph 
over $\langle\Sigma,\nodes,\vars\rangle$
such that:
\begin{itemize}
\item{} $\nodes_{g'} = \ren(\nodes_g)$,
\item{} $\xlabel_{g'}(\ren(n))=\xlabel_g(n)$, for all $n \in \nodes_g$,
\item{} $\xsucc_{g'}(\ren(n)) = \ren(n_1)\ren(n_2)\ldots \ren(n_k)$, \\
  iff $\xsucc_g(n) = n_1 n_2 \ldots n_k$,
  for all $n,n_1,\ldots n_k \in \nodes_g$,
\item{} $\Root_{g'} = \ren(\Root_g)$;   %%% changed from rta06
\item{} $\dom_{g'}(\ren(n))=\ren(\dom_g(n))$, for all non-root $n \in \nodes_g$.
\end{itemize}
%\vspace*{-4.5ex}
\end{definition}
In simpler language, $g'$ is
equal to $g$, except that the nodes in $\nodes_p$ have been renamed with a ``fresh'' name in $g'$.
In particular, the dominator of a renamed node $n$
is the renamed dominator of $n$.

Although intuitively simple, a formal definition
of the bubbling relation is non-trivial \cite{AntoyBrownChiang06Termgraph}.
To ease understanding, we split it into two parts.
Part I \cite{AntoyBrownChiang06RTA} defines the transformation
and implicitly defines the labeling and successor mapping
through the expression notation of graphs.
Part II, which is novel, defines the dominator mapping
which is not present in
\cite{AntoyBrownChiang06Termgraph,AntoyBrownChiang06RTA}
and cannot be inferred from the notation.
\begin{definition}[Bubbling \rm{- part I}]
\label{bubbling-definition-I}
Let $g$ be a graph and $c$ a node of $g$
such that the subgraph of $g$ at $c$
is of the form $x\,\mbox{\code{?}}\,y$,
i.e., $g|_c=x\,\mbox{\code{?}}\,y$.
Let $d$ be a dominator of $c$ in $g$ and
$\nodes_p$ the set of nodes that are on some path
from $d$ to $c$ in $g$, including $d$ and $c$,
i.e., $\nodes_p = \{n \mid n_1n_2\ldots n_k \in \Path_g(d,c)
\;\mbox{and $n=n_i$ for some $i$}\}$, where $\Path_g(d,c)$ is the
set of all paths from $d$ to $c$ in $g$.
%\todo{could this just be $\nodes_p = \bigcup \Path_g(d,c)$?}
%
Let $\ren_{x}$ and $\ren_{y}$ be partial renamings of $g$ 
with domain $\nodes_p$ and disjoint images.
Let $g_q = \ren_{q}(g|_d[c \leftarrow q])$, for $q \in \{x,y\}$.
The \emph{bubbling} relation on graphs is
denoted by ``$\simeq$'' and defined by
$g \simeq g[d \leftarrow g_x\mbox{\code{?}}\,g_y]$,
where the root node of the replacement of $g$ at $d$ is
obviously fresh.
We call $c$ and $d$ the \emph{origin} and \emph{destination},
respectively, of the bubbling step, and we denote the
step with ``$\simeq_{cd}$'' when this information is relevant.
\end{definition}
\begin{definition}[Bubbling \rm{- part II}]
  \label{bubbling-definition-II}
  Let $g$ and $g'$ be graphs such that $g \simeq_{cd} g'$,
  for some nodes $c$ and $d$ of $g$.
  For each node $n'$ of $g'$, we define $\dom_{g'}(n')$
  by cases as follows.
  \begin{itemize}
  \item [(1)]
    node $n'$ is created by the bubbling transformation.
    \begin{itemize}
    \item [(a)] $n'=r$, where $r$ is the root (labeled by the choice)
      of the replacement at $d$ in $g$: $\dom_{g'}(r) = \dom_{g}(d)$.
      % if $d$ (and $r$) are roots, that's OK
    \item [(b)] $n'=\ren(n)$ and $n' \ne n$,
      i.e., $n'$ properly renames some node $n$ of $g$
      according to some renaming $\ren$, i.e., $n'$ is a node in
      a path from $d$ down to the choice.  We consider two cases:
      \begin{itemize}
      \item [(b.1)] $n=d$, i.e., $n'$ renames the destination of the
        step and consequently is a successor of $r$: $\dom_{g'}(n') = r$.
        % i.e., the dominator of $n'$ is the root of the replacement;
      \item [(b.2)] $n \ne d$: $\dom_{g'}(n') = \ren(\dom_{g}(n))$.
        % i.e., the dominator of $n'$ is the renaming of the dominator of $n$.
      \end{itemize}
    \end{itemize}
  \item [(2)]
    node $n'$ persists from $g$ to $g'$:
    \begin{itemize}
    \item [(c)] let $n'=\ren(n)$ and $n' = n$. We consider three cases:
      \begin{itemize}
      \item [(c.1)] $n$ is the left (resp.~right) successor of $c$:
        $\dom_{g'}(n) = \dom_{g}(z)$, where $z$ left (resp.~right)
        instance of $\ren(c)$.
        % i.e., the left (resp. right) clone of $d$
      \item [(c.2)] $\dom_{g'}(n) \ne \ren(\dom_g(n))$:
        % i.e., $(\dom_g(n)$ was cloned, but the node wasn't.
        % Then we set $\dom_{g'}(n)$ to the root of the cloned expression,
        $\dom_{g'}(n') = r$.
      \item [(c.3)] $\dom_{g'}(n) = \ren(\dom_g(n))$:
        % i.e., $\dom_g(n)$ is not cloned
        $\dom_{g'}(n) = \dom_{g}(n)$.
      \end{itemize}
    \end{itemize}
  \end{itemize}
\end{definition}
This definition is sensible,
i.e., if $\dom_g$ is correct, then $\dom_{g'}$ is correct.
Later, we will provide an informal proof of this claim.
\begin{figure}[hbt]
  \begin{mdframed}
    \centerline{
      \xymatrix@C=-2pt@R=15pt@C=5pt{
        & & & & & & &\code{>} \ar@{-}[dlll] \ar@{-}[drr] \\
        & & & & \code{?} \ar@{-}[dlll] \ar@{-}[drr] & & & & & \code{23} \\
        & \code{/} \ar@{-}[dl] \ar@{-}[dr] & & & & & \code{/} \ar@{-}[dl] \ar@{-}[dr]\\
        \code{weight} \ar@{-}@/_9pt/[ddr] & & \code{\power} \ar@{-}[dl]
            \ar@{-}`d[r]`[rrrrrrd]    % \ar@{-}`d[r]`[rrrrrrd]
            & & & \code{weight} \ar@{-}@/_9pt/[ddr] & & \code{\power} \ar@{-}@/^5pt/[dr] \ar@{-}[dl]  \\
        & \code{height} \ar@{-}[d] & & & & & \code{height} \ar@{-}[d] & & \code{2} \\
        & \code{Alice} & & & & & \code{parent} \ar@{-}[d] \\
        & & & & & & \code{Bob} \\            
      }
    }
    \caption{
      \label{one-step}
      Result of executing a bubbling step on the state
      the computation, say $e$, of Figure \protect{\ref{bmi-graph}}.
      For any node $n$ of $e$ on a path from the dominator to \emph{choice},
      such as the node labeled by \emph{weight},
      the bubbling step creates two new nodes with the same label in the result.
      We call these nodes ``clones'' of $n$.
    }
  \end{mdframed}  
\end{figure}
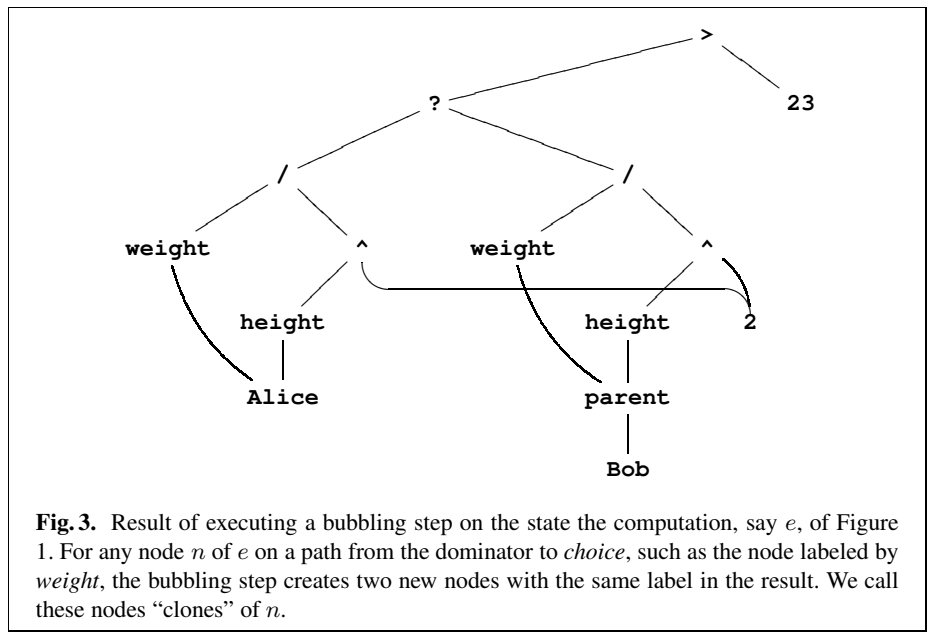

\section{Algorithms}
\label{Algorithms}

A non-deterministic step executed by bubbling must compute
a dominator, ideally the immediate one, of a node in a graph, $g$.
Computing the immediate dominator of a node in $g$
requires the traversal of $g$ \cite{LengauerTarjan79ACM}.
This requirement makes bubbling too expensive
when a non-deterministic step occurs in a large context.
To avoid this cost, we attach to each node $n$
of $g$ a dominator of $n$, ideally the immediate one,
and we maintain this attribute during a computation.
This attribute allows us to execute bubbling steps
using only readily available, local information.

\begin{figure}[htb]
  \begin{mdframed}
  \baselineskip=1.1\baselineskip
  \hspace*{1ex}
  \begin{minipage}[t]{0.455\textwidth}
bubble$(c)$ \\
\w let $r$ be a fresh node \\
\w $\xlabel(r):=\;$'\code{?}' \\
\w $d := \dom_g(c)$ \\
\w $\dom_{g'}(r) := \dom_g(d)$ \\
\w for $s$ in \{\emph{left, right}\}, do \\
\w\w clear map \\
\w\w traverse$(c,d)$ \\
\w\w add map$[d]$ to $\xsucc(r)$ \\
\w\w $\dom_{g'}($map$[d]) := r$ \\
\w\w replace map$[c]$ with $s(\xsucc(c))$ \\
\w replace $d$ with $r$
  \end{minipage}
  \begin{minipage}[t]{0.480\textwidth}
%%% Say traverse is nested inside bubble
traverse$(x,d)$ \\
\w if $x$ is mapped, then return \\
\w let $x'$ be a fresh node \\
\w map$[x]:=x'$ \\
\w $\xlabel(x'):=\xlabel(x)$ \\
\w if $x \ne d$, then \\
\w\w for $y$ in $\xsucc^{-1}(x)$, do \\
\w\w\w traverse$(y,d)$ \\
\w\w\w add $x'$ to $\xsucc({\rm map}[y])$ \\
\w\w $\dom_{g'}(x'):={\rm map}[\dom_g(x)]$ \\
\w for $z$ in $\dom_g^{-1}(x)$, do \\
\w\w if $z$ is not mapped, then $\dom_{g'}(z) := r$
  \end{minipage}
  \vspace*{.1ex}
  \caption{\label{bubbling-algorithm}
    Bubbling step algorithm.  This algorithm executes a
    bubbling step $g \simeq_{cd} g'$ that preserves the correctness
    of the dominator attribute.  An explanation is in the flowing text.
    The step visits only nodes in the path(s) from the source, $c$, to
    the destination, $d$.
  }
  \vspace*{.3ex}  
  \end{mdframed}
\end{figure}

The procedures in Fig.~\ref{bubbling-algorithm} execute bubbling steps.
\emph{Traverse}, which is in the scope of \emph{bubble},
executes a depth-first \emph{upward}
partial traversal of $g$, starting at some node $c$.
The procedure terminates upon either visiting a
node already visited, or visiting $d=\dom_g(c)$.
Since, by definition, $d$ is on every path from the root of $g$
to $c$, \emph{traverse} terminates for every node $c$.

An execution of \emph{traverse} clones each node on each path from $d$
to $c$.  A function, \emph{map}, maps each node being cloned to its clone.
\emph{Traverse} clones a node $x$ producing $x'$,
recursively clones all the paths from $d$ to each predecessor of $x$,
installs $x'$ as their successor, and sets label and dominators of $x'$.
If $x$ dominates a node $z$ which is not mapped,
and therefore is outside the portion
of $g$ being cloned by \emph{traverse}, then $x'$ will not be
the dominator of $z$ after the bubbling step.  This is because
a second clone of $x$ is created by another invocation of \emph{traverse}.
Thus, the dominator of $z$ is
set to node $r$, which is created by procedure \emph{bubble}, discussed next.
This situation is exemplified by the node labeled by $2$ in
Fig.~\ref{one-step}.

Procedure \emph{bubble} executes a bubbling step at some node $c$.
Its main activity, delegated to \emph{traverse},
is to clone the portion of $g$ between $c$ and $d=\dom_g(c)$ twice.
Let's call these clones $g_l$ and $g_r$.
The remaining activities are ``gluing''
together the pieces according to the definition of a bubbling step.
A fresh node, $r$, labeled by \ccode{?} is the predecessor
of $g_l$ and $g_r$.  The instruction \emph{replace p with q}
means that every incoming edge to $p$ is redirected
\cite[Def. 8]{EchahedJanodet97IMAG} to $q$
and that $\dom_{g'}(q)=\dom_g(p)$.
The bottom-most nodes of $g_l$ and $g_r$ are
the clones of $c$.  These nodes are discarded and replaced by the
left and right arguments of $c$.  Successor and dominator
attributes are adjusted as needed.

Our algorithm uses the inverses of both the dominator and
successor relations.  Hence, dominance should be implemented bidirectionally.
In order for the bubbling algorithm just described to work, we
must maintain the dominator attribute of a graph as rewriting
steps are executed.  The initial top-level expression
typically has only a few nodes; sometimes it is just ``\emph{main}.''
For this reason initializing the attribute is trivial.

When a rewrite step $e \to e'$ is executed, for some graphs $e$ and
$e'$, some nodes of $e$ are missing in $e'$ and typically some 
nodes that were not in $e$ appear in $e'$.
A consequence is that the $\dom$ attribute of some nodes in $e'$
may be missing or may be incorrect.
For this reason, after executing a rewrite step,
it is required to set the $\dom$ attribute of these nodes.
This is achieved by executing the procedure \emph{fix\_dominator}.

The notion of \emph{redex} is crucial in the following discussion,
but how to determine the redex of step is not a part of the discussion.
Choosing the redex of step is the job of an evaluation
strategy, but our algorithm is independent of the strategy.

Let $g$ be a graph, $f$ be the root of the redex, and
$l \to r $ be the rule of the step.  Node $f$ is not labeled by
\emph{choice}, otherwise we would execute a bubbling step.
The step replaces the redex at $f$ with a replacement rooted by
some node $e$.  The \emph{redex pattern} \cite[Def. 2.7.3]{Terese03} at $f$ 
consists of the set of nodes of $g$ matched by non-variable nodes of $l$.
Any node in the set is erased unless it is shared
(there is an edge incoming to the node from a node outside the redex pattern).
By analogy, we call the set of nodes created by the step the 
\emph{contractum pattern} (even though it is not a pattern).
Any node in the set corresponds to a non-variable node of $r$.
The dominator-preserving rewriting algorithm is presented
in Fig~\ref{preserving-rewriting}.

\newcommand{\sml}[1]{{\phantom{x}\small #1.~~}}
\begin{figure}[htb]
\begin{mdframed}
\baselineskip=1.1\baselineskip
\hspace*{1ex}
\begin{minipage}[t]{.95\textwidth}
fix\_dominator$(f)$ \\
\sml0 let $e$ be the root of the contractum; \\
\sml1 for each node $d$ in the redex pattern at $f$, erased by the step,
      $\dom_{g'}(\dom_g^{-1}(d)) := e$; \\
\sml2 for each node $c$ in the contractum pattern at $e$, excluded $e$, $\dom_{g'}(c) := e$; \\
\sml3 $\dom_{g'}(e) := \dom_g(f)$; \\
\sml4 for every other node $n$ of $g'$ (and $g$), $\dom_{g'}(n) := \dom_g(n)$;
\end{minipage}
\caption{\label{preserving-rewriting}
  Dominator preserving step algorithm.  After a rewrite step
  $g \to g'$ at $f$, the algorithm associates a dominator to the nodes
  affected by the step.  Any node dominated by a node that is
  erased by the step becomes dominated by the root of the contractum.
  Any node created by the step becomes dominated by the root of the contractum.
  The root of the contractum becomes dominated by the dominator of the
  root of the redex.  The dominator of any other node is unchanged.  }
\end{mdframed}
\end{figure}
Procedure \emph{fix\_dominator} sets or updates the $\dom$ attribute
of nodes that are created or nodes whose dominator is erased by a
rewrite step, respectively.
For all other nodes,
if the $\dom$ attribute is correct before the rewrite step,
then it remains correct after the rewrite step.
Later, we will provide an informal proof of this claim.

Procedure \emph{fix\_dominator} may not always produce the \emph{immediate}
dominator of some node. Consider the following rule and expression:
\equprogram{%
  f x y = h y \\
  f (g z) (g z) where z = 0
}
where $f$, $g$ and $h$ are operations and $x$, $y$ and $z$ variables.
The expression rewrites to \code{h\;(g\;0)}. The dominator of \code{0}
is set, by our algorithm,
to \code{h} whereas the immediate dominator is \code{g}.
Since contractum
patterns are typically shallow and with few nodes, the
dominator set by the algorithm for any node $n$
will typically be rather close to $n$.

\section{Correctness}
\label{Correctness}

The correctness of our algorithm is stated in
propositions \ref{rewriting} and \ref{prop-bubble}.
We will present each proposition and sketch a short proof idea.
This is not meant to be a formal, rigorous proof,
but rather to give some intuition on why these algorithms
perform as intended.

Both of these proofs rely on more elementary facts about dominators.
First, removing edges from a graph does not change the dominance relation
as long as the graph remains connected.
Second, if $d$ dominates $c$, then adding the edge $(s,t)$ to a graph
rooted by $r$ will remove a dominator 
iff there is a path $r \ldots s, t \ldots c$ that does not contain $d$.
Third, if $d$ dominates $c$, then $d$ dominates every node on
every path from $d$ to $c$.
The first proposition states that a rewriting
followed by the execution of procedure $\emph{fix\_dominator}$
preserves the correctness of the dominator attribute.

\begin{proposition}[Correctness of fix\_dominator]
  \label{rewriting}
  Let $g$ be a graph such that the dominator attribute of $g$ is
  correct and $f$ be a redex of $g$.
  If $g'$ is the graph obtained from $g$ by first executing a rewrite at $f$,
  and then executing \emph{fix\_dominator}$(f)$,
  then the dominator attribute of $g'$ is correct.
\end{proposition}

\noindent
Let $n$ be an arbitrary node of $g'$.
The proof is by cases on the relative positions of $n$, $\dom_g(n)$, and $f$.
There are 4 cases to consider.
If $\dom_{g}(n)$ is erased by the rewrite, then the root of the contractum
is a dominator of $n$ in $g'$.
If $n$ is the root of the contractum,
then $\dom_{g}(f)$ is a dominator of $n$ in $g'$.
If $n$ is any other node in the contractum pattern,
then the root of the contractum
is a dominator of $n$ in $g'$.
If $n$ is any other node,
then $\dom_g(n)$ is a dominator of $n$ in $g'$.
This last case follows from the fact that,
if $\dom_g(n)$ does not dominate $f$,
then the dominance relation is trivially preserved,
and if it does dominate $f$,
then it will dominate the contractum,
and therefore will dominate $c$.
\\[1ex]
The second proposition states that procedure \emph{bubble}
preserves the correctness of the dominator attribute.
%%% it should also show that it computes a bubbling step.

\begin{proposition}[Correctness of bubble]
  \label{prop-bubble}
  Let $g$ be a graph such that the
  dominator attribute of $g$ is correct and $c$ is a node of $g$
  labeled by \ccode{?}.  Let $g'$ be the graph obtained from $g$ by
  executing bubble($c$).  Then the dominator attribute of $g'$ is
  correct.
\end{proposition}

\noindent
Let $P$ be the set of nodes on a path from $c$ to $\dom_g(c)$.
Let $\ren_x$, where $x$ is either \emph{left} or \emph{right},
be a renaming created in the bubbling step.
The bubbling step creates two subgraphs
$P_{\emph{left}}$ and $P_{\emph{right}}$ isomorphic to $P$.
Let $r$ be the predecessor in $g'$ of the roots
of $P_{\emph{left}}$ and $P_{\emph{right}}$.
As in Prop.~\ref{rewriting}, we show the correctness by cases.
Let $n$ be an arbitrary node in $g'$.
If $n=c$, the source of the bubbling step,
$\dom_g(\dom_g(c))$ is a dominator of $n$ in $g'$.
If $n=\ren_x(\dom_g(c))$, then $r$ is a dominator of $n$ in $g'$
since $r$ is the only predecessor of $n$.
If both $n$ and $\dom_g(n)$ are in $P$,
then the dominator of the renaming of $n$ in $g'$ is
the renaming of the dominator of $n$ in $g$.
That is, $\dom_{g'}{\ren_x(n)} = \ren_x(\dom_{g}(n))$.
This follows directly from the fact that $P_x$ is isomorphic to $P$.
If $n\not\in P$, but $\dom_g(n)\in P$,
then $r$ is a dominator of $n$ in $g'$,
since the bubbling steps creates two ``copies'' of $\dom_g(n)$.
If neither $n$ nor $\dom_g(n)$ are in $P$, then $\dom_{g}(n)$
is a dominator of $n$ in $g'$.
This final case is similar to the final case of the previous proposition.
If $\dom_g(n)$ doesn't dominate $\dom_g(c)$
then the dominance relation is trivially preserved, but if it does
dominate $\dom_g(c)$ then it will dominate $r$,
so the dominance relation will still be preserved.

\section{Related Work}
\label{Related Work}

Several techniques have been proposed to execute
non-deterministic steps in functional logic programming.
\emph{Backtracking} \cite{knuth}
arbitrarily chooses either alternative of a \emph{choice}.
If the computation of the chosen alternative fails to find a value,
the computation continues with the other alternative.
Otherwise, if and when the computation of the chosen alternative completes,
the result is presented to the user and
the computation continues with the other alternative.
\emph{Copying} is a naive approach to the problem
of a single context for the two alternatives of a \emph{choice}.
It ``moves'' the \emph{choice} to the root
of the expression being evaluated by cloning
the path(s) from the root down to the \emph{choice}.
In one clone, the \emph{choice} is replaced by one of its alternatives.
In the other clone it is replaced by the other alternative.
The two clones are evaluated independently and simultaneously,
e.g., by means of interleaved steps.
Copying is not used in practice,
but is refined by bubbling and pull-tabbing described next.

\emph{Bubbling} \cite{AntoyBrownChiang06Termgraph,AntoyBrownChiang06RTA}
attempts to improve the efficiency of copying by ``moving''
the \emph{choice} to a dominator of the \emph{choice}.
Only the paths between the dominator and the \emph{choice} are cloned.
The upfront number of cloned nodes is reduced.
If the computation of one clone fails, the \emph{choice} ``disappears'' and
the overall amount of cloning is reduced.
\emph{Pull-tabbing} \cite{Antoy11ICLP,Brassel2011PHD,BrasselHuchAPLAS07}
attempts to improve the efficiency of bubbling by ``moving''
the \emph{choice} to a predecessor of the \emph{choice}.
There is no path to clone, but in some cases the step would be unsound.
For this reason an additional piece of information,
called a \emph{fingerprint}, is attached to nodes
to ensure that subexpressions resulting from one alternative of a \emph{choice}
are not mixed with subexpressions resulting from the other
alternative of the same \emph{choice}.

All the above techniques have drawbacks---some serious.
Backtracking may fail to produce the results of
the second alternative of a \emph{choice} if the computation
of the first alternative does not terminate.
Copying may needlessly clone a long path that remains
largely unused if either alternative of a \emph{choice} quickly fails.
Bubbling, before our formulation,
required traversing the entire expression being evaluated
to find a suitable dominator of the \emph{choice} \cite{LengauerTarjan79ACM}.
Pull-tabbing has a substantial overhead and must bring
every \emph{choice} to the top of an expression, even after an alternative
of the \emph{choice} fails.

The usual application of dominators is to optimizing
control flow graphs, which are static.
Bubbling and rewriting keep updating the expression being evaluated.
Hence, our work explores novel applications of graph dominance.

\section{Conclusion}
\label{Conclusion}

We propose an original algorithm that executes bubbling steps accessing only
local information.  To this aim we add a dominator attribute to the
graph's nodes.  We also extend the notions of rewriting and bubbling
for maintaining the attribute during a computation.  The attribute
allows us to execute a step with an effort independent of the size of the
context in which the step occurs. 

We roughly estimate the overhead of our approach as follows.
Most steps of a computation are deterministic, hence rewrites.
The cost of a rewrite is that of pattern matching and
executing replacements.
Pattern matching is unaffected by our extensions. 
Replacements consist in allocating nodes and setting
their attributes.
A drawback of our approach is the presence of additional attributes
in the nodes, which must be set.  In plausible
representations of expressions, regardless of our approach,
a node has several attributes such as label and successors.
We add dominator and predecessors.
A gross estimate is that our approach doubles
the number of attributes that must be allocated and set.
Processing the attributes of a node is only a fraction of the work
of rewriting, hence we estimate that the overhead
of maintaining the dominator is a factor less than 2
both for memory allocation and execution time.

Our work suggests that a compiler of a functional logic language, whose
backend generates bubbling code, is not only feasible, but possibly competitive.
Bubbling has some unique advantages over other techniques for the
implementation of non-determinism: completeness w.r.t.{} backtracking,
reduced cloning w.r.t.{} copying, and no overhead for failures
w.r.t.{} pull-tabbing.
Future work should aim at a more formal and detailed
proof of the correctness of our
algorithms and at an implementation to assess their competitiveness.

\bibliographystyle{plain}

\end{document}